\documentclass[12pt]{article}
\usepackage{amssymb,amsmath,epsfig,graphicx}
\begin{document}
\title{\bf Bianchi Type $I$ Cosmology in $f(R,T)$ Gravity}

\author{M. Farasat
Shamir\thanks{farasat.shamir@nu.edu.pk}\\\\  Department of
Sciences \& Humanities, \\National University of Computer \&
Emerging Sciences,\\ Lahore Campus, Pakistan.}

\date{}

\maketitle
\begin{abstract}
This paper is devoted to investigate the exact solutions of
Bianchi type $I$ spacetime in the context of $f(R,T)$ gravity
\cite{fRT1}, where $f(R,T)$ is an arbitrary function of Ricci scalar $R$
and trace of the energy momentum tensor $T$.
For this purpose, we find two exact solutions using the
assumption of constant deceleration parameter and
variation law of Hubble parameter. The obtained solutions
correspond to two different models of the universe. The physical
behavior of these models is also discussed.
\end{abstract}

{\bf Keywords:} $f(R,T)$ gravity, Bianchi type $I$, deceleration parameter.\\
{\bf PACS:} 04.50.Kd.

\section{Introduction}

The most popular issue in the modern day cosmology is the current
expansion of universe. It is now evident from observational and
theoretical facts that our universe is in the phase of accelerated
expansion \cite{acc1}-\cite{acc9}. The phenomenon of dark energy and
dark matter is another topic of discussion \cite{de1}-\cite{de8}. It
was Einstein who first gave the concept of dark energy and
introduced the small positive cosmological constant. But after sometime,
he remarked it as the biggest mistake in his life. However, it is
now thought that the cosmological constant may be a suitable
candidate for dark energy. Another proposal to justify the current
expansion of universe comes from modified or alternative theories of
gravity. $f(T)$ theory of gravity is one such example which has been
recently developed. This theory is a generalized version of
teleparallel gravity in which Weitzenb\"{o}ck connection is used
instead of Levi-Civita connection. The interesting feature of the
theory is that it may explain the current acceleration without
involving dark energy. A considerable amount of work has been done
in this theory so far \cite{ft}. Another interesting modified theory
is $f(R)$ theory of gravity which involves a general function of
Ricci scalar in standard Einstein-Hilbert Lagrangian. Some
review articles \cite{rev} can be helpful to understand the theory.

Many authors have investigated $f(R)$ gravity in different
contexts \cite{Analysis of $f(R)$ Theory Corresponding to NADE and
NHDE}-\cite{f(R)gravity constrained by PPN parameters and
stochastic background of gravitational waves}. Spherically
symmetric solutions are most commonly studied solutions
due to their closeness to the nature. Multam$\ddot{a}$ki and Vilja
\cite{fr1} explored vacuum and perfect fluid solutions of
spherically symmetric spacetime in metric version of this theory.
They used the assumption of constant scalar curvature and found
that the solutions corresponded to the already existing solutions
in general relativity (GR). Noether symmetries have been used by
Capozziello et al. \cite{fr2} to study spherically symmetric
solutions in $f(R)$ gravity. Similarly many interesting results
have been found using spherical symmetry in $f(R)$ gravity
\cite{fr3}. Cylindrically symmetric vacuum and non-vacuum
solutions have also been explored in this theory \cite{cylndr}.
Sharif and Shamir \cite{me1} found plane symmetric solutions. The
same authors \cite{me2} discussed the solutions of Bianchi types
$I$ and $V$ cosmologies for vacuum and non-vacuum cases. Conserved
quantities in $f(R)$ gravity via Noether symmetry approach have
been recently calculated \cite{me3}.

In a recent paper \cite{fRT1}, Harko et al. proposed a new
generalized theory known as $f(R,T)$ gravity. In this theory,
gravitational Lagrangian involves an arbitrary function of the
scalar curvature $R$ and the trace of the energy momentum tensor $T
$. Myrzakulov \cite{fRT2} discussed $f(R,T)$ gravity in which he
explicitly presented point like Lagrangians. Sharif and
Zubair \cite{fRT5} studied the laws of thermodynamics in this
theory. The same authors \cite{fRT45} investigated holographic and
agegraphic $f(R,T)$ models. Houndjo \cite{fRT4} reconstructed $f(R,T)$
gravity by taking $f(R,T)=f_1(R)+f_2(T)$ and it was proved that $f(R,T)$ gravity
allowed transition of matter from dominated phase to an acceleration
phase. Thus it is hoped that $f(R,T)$ gravity may explain the resent
phase of cosmic acceleration of our universe. This theory can be
used to explore many issues and may provide some satisfactory
results.

The isotropic models are considered to be most suitable to study
large scale structure of the universe. However, it is believed
that the early universe may not have been exactly uniform. This
prediction motivates us to describe the early stages of the
universe with the models having anisotropic background. Thus,
the existence of anisotropy in early phases of the universe is an interesting
phenomenon to investigate. A Bianchi Type $I$ cosmological model, being the generalization of flat
Friedmann-Robertson-Walker (FRW) model, is one of the simplest
models of the anisotropic universe. Therefore, it
seems interesting to explore Bianchi
type models in the context of $f(R,T)$ gravity. Adhav \cite{fRT3} investigated
the exact solutions of $f(R,T)$ field equations for locally rotationally symmetric Bianchi
type $I$ spacetime. Reddy et al. \cite{fRT46}
explored the solutions of Bianchi type $III$ spacetime
using the law of variation of Hubble's parameter. Bianchi type $III$ dark
energy model in the presence of perfect fluid source has been reported \cite{fRT47}.
Ahmed and Pradhan \cite{fRT48} studied Bianchi Type $V$ cosmology in this theory
by involving the cosmological constant in the field equations. Naidu et al. \cite{fRT49} gave the
solutions of Bianchi type $V$ bulk viscous string cosmological model.

In this paper, we are focussed to investigate the exact solutions
of Bianchi type $I$ spacetime in the framework of $f(R,T)$
gravity. The plan of paper is as follows: In section \textbf{2},
we give some basics of $f(R,T)$ gravity. Section \textbf{3} provides
the exact solutions for Bianchi type $I$ spacetime. Concluding
remarks are given in the last section.

\section{Some Basics of $f(R,T)$ Gravity}

The $f(R,T)$ theory of gravity is the generalization or
modification of GR. The action for this theory is given by
\cite{fRT1}
\begin{equation}\label{frt1}
S=\int\sqrt{-g}\bigg(\frac{1}{16\pi{G}}f(R,T)+L_{m}\bigg)d^4x,
\end{equation}
where $f(R,T)$ is an arbitrary function of the Ricci scalar $R$
and the trace $T$ of energy momentum tensor $T_{\mu\nu}$ while
$L_{m}$ is the usual matter Lagrangian. It is worth mentioning
that if we replace $f(R,T)$ with $f(R)$, we get the action for
$f(R)$ gravity and replacement of $f(R,T)$ with $R$ leads to the
action of GR. The energy momentum tensor $T_{\mu\nu}$ is defined
as \cite{emt}
\begin{equation}\label{frt2}
T_{\mu\nu}=-\frac{2}{\sqrt{-g}}\frac{\delta(\sqrt{-g}L_m)}{\delta
g^{\mu\nu}}.
\end{equation}
Here we assume that the dependance of matter Lagrangian is merely
on the metric tensor $g_{\mu\nu}$ rather than its derivatives. In
this case, we get
\begin{equation}\label{frt3}
T_{\mu\nu}=L_m g_{\mu\nu}-2\frac{\delta L_m}{\delta g^{\mu\nu}}.
\end{equation}
The $f(R,T)$ gravity field equations are obtained by varying the
action $S$ in Eq.(\ref{frt1}) with respect to the metric tensor
$g_{\mu\nu}$
\begin{equation}\label{frt4}
f_R(R,T)R_{\mu\nu}-\frac{1}{2}f(R,T)g_{\mu\nu}-(\nabla_{\mu}
\nabla_{\nu}-g_{\mu\nu}\Box)f_R(R,T)=\kappa
T_{\mu\nu}-f_T(R,T)(T_{\mu\nu}+\Theta_{\mu\nu}),
\end{equation}
where $\nabla_{\mu}$ denotes the covariant derivative and
\begin{equation*}
\Box\equiv\nabla^{\mu}\nabla_{\mu},~~ f_R(R,T)=\frac{\partial
f_R(R,T)}{\partial R},~~ f_T(R,T)=\frac{\partial
f_R(R,T)}{\partial
T},~~\Theta_{\mu\nu}=g^{\alpha\beta}\frac{\delta
T_{\alpha\beta}}{\delta g^{\mu\nu}}.
\end{equation*}
Contraction of (\ref{frt4}) yields
\begin{equation}\label{frt04}
f_R(R,T)R+3\Box f_R(R,T)-2f(R,T)=\kappa T-f_T(R,T)(T+\Theta),
\end{equation}
where $\Theta={\Theta_\mu}^\mu$. This is an important equation
because it provides a relationship between Ricci scalar $R$ and
the trace $T$ of energy momentum tensor. Using matter Lagrangian
$L_m$, the standard matter energy-momentum tensor is derived as
\begin{equation}\label{frt5}
T_{\mu\nu}=(\rho + p)u_\mu u_\nu-pg_{\mu\nu},
\end{equation}
where $u_\mu=\sqrt{g_{00}}(1,0,0,0)$ is the four-velocity in
co-moving coordinates and $\rho$ and $p$ denote energy density
and pressure of the fluid respectively. Perfect fluids problems
involving energy density and pressure are not any easy task to
deal with. Moreover, there does not exist any unique definition
for matter Lagrangian. Thus we can assume the matter Lagrangian as
$L_m=-p$ which gives
\begin{equation}\label{frt6}
\Theta_{\mu\nu}=-pg_{\mu\nu}-2T_{\mu\nu},
\end{equation}
and consequently the field equations (\ref{frt4}) take the form
\begin{equation}\label{frt7}
f_R(R,T)R_{\mu\nu}-\frac{1}{2}f(R,T)g_{\mu\nu}-(\nabla_{\mu}
\nabla_{\nu}-g_{\mu\nu}\Box)f_R(R,T)=\kappa
T_{\mu\nu}+f_T(R,T)(T_{\mu\nu}+pg_{\mu\nu}),
\end{equation}
It is mentioned here that these field equations depend on the
physical nature of matter field. Many theoretical models
corresponding to different matter contributions for $f(R,T)$
gravity are possible. However, Harko et al. \cite{fRT1} gave three classes of
these models
\[ f(R,T)= \left\lbrace
  \begin{array}{c l}
    {R+2f(T),}\\
    {f_1(R)+f_2(T),}\\{f_1(R)+f_2(R)f_3(T).}
  \end{array}
\right. \]\\
In this paper we are focussed to the first class, i.e.
$f(R,T)=R+2f(T)$. For this model the field equations become
\begin{equation}\label{frt8}
R_{\mu\nu}-\frac{1}{2}Rg_{\mu\nu}=\kappa
T_{\mu\nu}+2f'(T)T_{\mu\nu}+\bigg[f(T)+2pf'(T)\bigg]g_{\mu\nu},
\end{equation}
where prime represents derivative with respect to $T$.

\section{Exact Solutions of Bianchi Type $I$ Universe}

In this section, we shall find exact solutions of Bianchi I
spacetime in $f(R,T)$ gravity. For the sake of simplicity, we use
natural system of units $(G=c=1)$ and $f(T)=\lambda T$, where
$\lambda$ is an arbitrary constant. For Bianchi type $I$
spacetime, the line element is given by
\begin{equation}\label{6}
ds^{2}=dt^2-A^2(t)dx^2-B^2(t)dy^2-C^2(t)dz^2,
\end{equation}
where $A,~B$ and $C$ are defined as cosmic scale factors. The
Bianchi $I$ Ricci scalar turns out to be
\begin{equation}\label{7}
R=-2\bigg[\frac{\ddot{A}}{A}+\frac{\ddot{B}}{B}+\frac{\ddot{C}}{C}
+\frac{\dot{A}\dot{B}}{AB}+\frac{\dot{B}\dot{C}}{BC}+\frac{\dot{C}\dot{A}}{CA}\bigg],
\end{equation}
where dot denotes derivative with respect to $t$.

Using Eq.(\ref{frt8}), we get four independent field equations,
\begin{eqnarray} \label{11}
\frac{\dot{A}\dot{B}}{AB}+\frac{\dot{B}\dot{C}}{BC}+
\frac{\dot{C}\dot{A}}{CA}=(8\pi+3\lambda)\rho-\lambda
p,\\\label{12} \frac{\ddot{B}}{B}+\frac{\ddot{C}}{C}
+\frac{\dot{B}\dot{C}}{BC}=\lambda
\rho-(8\pi+3\lambda)p,\\\label{13}
\frac{\ddot{C}}{C}+\frac{\ddot{A}}{A}
+\frac{\dot{C}\dot{A}}{AC}=\lambda
\rho-(8\pi+3\lambda)p,\\\label{14}
\frac{\ddot{A}}{A}+\frac{\ddot{B}}{B}
+\frac{\dot{A}\dot{B}}{AB}=\lambda \rho-(8\pi+3\lambda)p.
\end{eqnarray}
These are four non-linear differential equations with five unknowns
namely $A,~B,~C$, $\rho$ and $p$. Subtracting Eq.(\ref{13}) from
Eq.(\ref{12}), Eq.(\ref{14}) from Eq.(\ref{13}) and Eq.(\ref{14})
from Eq.(\ref{11}), we get respectively
\begin{eqnarray}\label{015}
\frac{\ddot{A}}{A}-\frac{\ddot{B}}{B}
+\frac{\dot{C}}{C}\bigg(\frac{\dot{A}}{A}-\frac{\dot{B}}{B}\bigg)=0,\\\label{016}
\frac{\ddot{B}}{B}-\frac{\ddot{C}}{C}
+\frac{\dot{A}}{A}\bigg(\frac{\dot{B}}{B}-\frac{\dot{C}}{C}\bigg)=0,\\\label{017}
\frac{\ddot{A}}{A}-\frac{\ddot{C}}{C}
+\frac{\dot{B}}{B}\bigg(\frac{\dot{A}}{A}-\frac{\dot{C}}{C}\bigg)=0.
\end{eqnarray}
These equations imply that
\begin{eqnarray}\label{15}
\frac{B}{A}=d_1\exp\bigg[{c_1\int\frac{dt}{a^3}}\bigg],\\\label{16}
\frac{C}{B}=d_2\exp\bigg[{c_2\int\frac{dt}{a^3}}\bigg],\\\label{17}
\frac{A}{C}=d_3\exp\bigg[{c_3\int\frac{dt}{a^3}}\bigg],
\end{eqnarray}
where $c_1,~c_2,~c_3$ and $d_1,~d_2,~d_3$ are integration
constants which satisfy the following relation
\begin{equation}\label{18}
c_1+c_2+c_3=0,\quad d_1d_2d_3=1.
\end{equation}
Using Eqs.(\ref{15})-(\ref{17}), we can write the unknown metric
functions in an explicit way
\begin{eqnarray}\label{19}
A=ap_1\exp\bigg[{q_1\int\frac{dt}{a^3}}\bigg],\\\label{20}
B=ap_2\exp\bigg[{q_2\int\frac{dt}{a^3}}\bigg],\\\label{21}
C=ap_3\exp\bigg[{q_3\int\frac{dt}{a^3}}\bigg],
\end{eqnarray}
where
\begin{equation}\label{22}
p_1=({d_1}^{-2}{d_2}^{-1})^{\frac{1}{3}},\quad
p_2=(d_1{d_2}^{-1})^{\frac{1}{3}},\quad
p_3=(d_1{d_2}^2)^{\frac{1}{3}}
\end{equation}
and
\begin{equation}\label{23}
q_1=-\frac{2c_1+c_2}{3},\quad q_2=\frac{c_1-c_2}{3},\quad
q_3=\frac{c_1+2c_2}{3}.
\end{equation}
It is mentioned here that $p_1,~p_2,~p_3$ and $q_1,~q_2,~q_3$ also
satisfy the relation
\begin{equation}\label{24}
p_1p_2p_3=1,\quad q_1+q_2+q_3=0.
\end{equation}

\subsection{Some Important Physical Parameters}

Now we present some important definitions of physical parameters.
The average scale factor $a$ and volume scale factor $V$ are defined
as
\begin{equation}\label{8}
a=\sqrt[3]{ABC}, \quad V=a^3=ABC.
\end{equation}
The generalized mean Hubble parameter $H$ is given by
\begin{equation}\label{008}
H=\frac{1}{3}(H_1+H_2+H_3),
\end{equation}
where
$H_1=\frac{\dot{A}}{A},~H_2=\frac{\dot{B}}{B},~H_3=\frac{\dot{C}}{C}$
are defined as the directional Hubble parameters in the directions
of $x,~y$ and $z$ axis respectively. The mean anisotropy parameter
$A$ is
\begin{equation}\label{0000009}
A=\frac{1}{3}\sum^3_{i=1}\bigg(\frac{H_i-H}{H}\bigg)^2.
\end{equation}
The expansion scalar $\theta$ and shear scalar $\sigma^2$ are
defined as follows
\begin{eqnarray}\label{09}
\theta&=&u^\mu_{;\mu}=\frac{\dot{A}}{A}+\frac{\dot{B}}{B}+\frac{\dot{C}}{C},\\
\label{00009} \sigma^2&=&\frac{1}{2}\sigma_{\mu\nu}\sigma^{\mu\nu}
=\frac{1}{3}\bigg[\bigg(\frac{\dot{A}}{A}\bigg)^2+\bigg(\frac{\dot{B}}{B}\bigg)^2
+\bigg(\frac{\dot{C}}{C}\bigg)^2-\frac{\dot{A}\dot{B}}{AB}-\frac{\dot{B}\dot{C}}{BC}
-\frac{\dot{C}\dot{A}}{CA}\bigg],~~
\end{eqnarray}
where
\begin{equation}\label{009}
\sigma_{\mu\nu}=\frac{1}{2}(u_{\mu;\alpha}h^\alpha_\nu+u_{\nu;\alpha}h^\alpha_\mu)
-\frac{1}{3}\theta h_{\mu\nu}
\end{equation}
with $h_{\mu\nu}=g_{\mu\nu}-u_{\mu}u_{\nu}$ defined as the projection tensor.
The deceleration parameter $q$ is the measure of the cosmic
accelerated expansion of the universe. It is defined as
\begin{equation}\label{26}
q=-\frac{\ddot{a}a}{\dot{a}^2}.
\end{equation}
The behavior of the universe models is determined by the sign of
$q$. The positive value of deceleration parameter suggests a
decelerating model while the negative value indicates inflation.
Since there are four equations (\ref{11})-(\ref{14}) and five unknowns,
so we need an additional constraint to solve them. Here we use a
well-known relation \cite{15} between the average scale factor $a$
and Hubble parameter $H$ to solve the equations,
\begin{equation}\label{27}
H=la^{-n},
\end{equation}
where $l$ and $n$ are positive constants. This is an important relation
because it yields a constant value of deceleration parameter and consequently we obtain power
law and exponential models of universe. Using Eqs.(\ref{008}) and (\ref{27}), we get
\begin{equation}\label{28}
\dot{a}=la^{1-n}
\end{equation}
and the deceleration parameter becomes
\begin{equation}\label{29}
q=n-1.
\end{equation}
Integrating Eq.(\ref{28}), it follows that
\begin{equation}\label{30}
a=(nlt+k_1)^{\frac{1}{n}},\quad n\neq0,
\end{equation}
and
\begin{equation}\label{31}
a=k_2\exp(lt),\quad ~~~n=0,
\end{equation}
where $k_1$ and $k_2$ are constants of integration.

\subsection{Singular Model of the Universe}

Here we investigate the model of universe when $n\neq0$, i.e.,
$a=(nlt+k_1)^{\frac{1}{n}}$. In this case, the metric coefficients
$A,~B$ and $C$ take the form
\begin{eqnarray}\label{35}
A&=&p_1(nlt+k_1)^{\frac{1}{n}}\exp\bigg[\frac{q_1(nlt+k_1)^
{\frac{n-3}{n}}}{l(n-3)}\bigg],\quad n\neq3,\\\label{36}
B&=&p_2(nlt+k_1)^{\frac{1}{n}}\exp\bigg[\frac{q_2(nlt+k_1)^
{\frac{n-3}{n}}}{l(n-3)}\bigg],\quad n\neq3,\\\label{37}
C&=&p_3(nlt+k_1)^{\frac{1}{n}}\exp\bigg[\frac{q_3(nlt+k_1)^
{\frac{n-3}{n}}}{l(n-3)}\bigg],\quad n\neq3.
\end{eqnarray}
The directional Hubble parameters $H_i$ ($i=1,2,3$) turn out to be
\begin{equation}\label{38}
H_i=\frac{l}{nlt+k_1}+\frac{q_i}{(nlt+k_1)^{\frac{3}{n}}}.
\end{equation}
The mean generalized Hubble parameter and volume scale factor are
\begin{equation}\label{39}
H=\frac{l}{nlt+k_1},\quad V=(nlt+k_1)^\frac{3}{n}.
\end{equation}
The mean anisotropy parameter becomes
\begin{equation}\label{3929}
A=\frac{{q_1}^2+{q_2}^2+{q_3}^2}{3l^2(nlt+k_1)^{(6-2n)/n}}.
\end{equation}
The expansion scalar and shear scalar for this model are given by
\begin{equation}
\theta=\frac{3l}{nlt+k_1},\quad\sigma^2=\frac{{q_1}^2+{q_2}^2+{q_3}^2}{2(nlt+k_1)^{6/n}}.
\end{equation}
Using Eqs. (\ref{11})-(\ref{14}), the energy density of the
universe turns out to be
\begin{eqnarray}\nonumber
\rho=&&\frac{1}{12(\lambda+2\pi)(\lambda+4\pi)}\bigg[4(\lambda+3\pi)
\bigg\{\frac{3l^2}{(nlt+k_1)^2}+\frac{q_1q_2+q_2q_3+q_3q_1}{(nlt+k_1)^{\frac{6}{n}}}\bigg\}\\\label{ro1}
&-&\lambda
\bigg\{\frac{3l^2(1-n)}{(nlt+k_1)^2}+\frac{{q_1}^2+{q_2}^2+{q_3}^2}{(nlt+k_1)^{\frac{6}{n}}}\bigg\}\bigg]
\end{eqnarray}
while the pressure of the universe becomes
\begin{eqnarray}\nonumber
p=&&\frac{-1}{12(\lambda+2\pi)(\lambda+4\pi)}\bigg[4\pi
\bigg\{\frac{3l^2}{(nlt+k_1)^2}+\frac{q_1q_2+q_2q_3+q_3q_1}{(nlt+k_1)^{\frac{6}{n}}}\bigg\}\\\label{p1}
&+&(3\lambda+8\pi)\bigg\{\frac{3l^2(1-n)}{(nlt+k_1)^2}+\frac{{q_1}^2+{q_2}^2+{q_3}^2}{(nlt+k_1)^{\frac{6}{n}}}\bigg\}\bigg].
\end{eqnarray}
The plots of $\rho$, $p$ and equation of state parameter $w=p/\rho$ against
time coordinate $t$ are shown in figure $1$ and $2$ respectively.
It is evident from figure $2$ that $w\rightarrow \frac{1}{3}$ as $t\rightarrow\infty$.
Thus the model corresponds to a radiation dominated universe as the time grows.
\begin{figure}\center
\begin{tabular}{cccc}
\epsfig{file=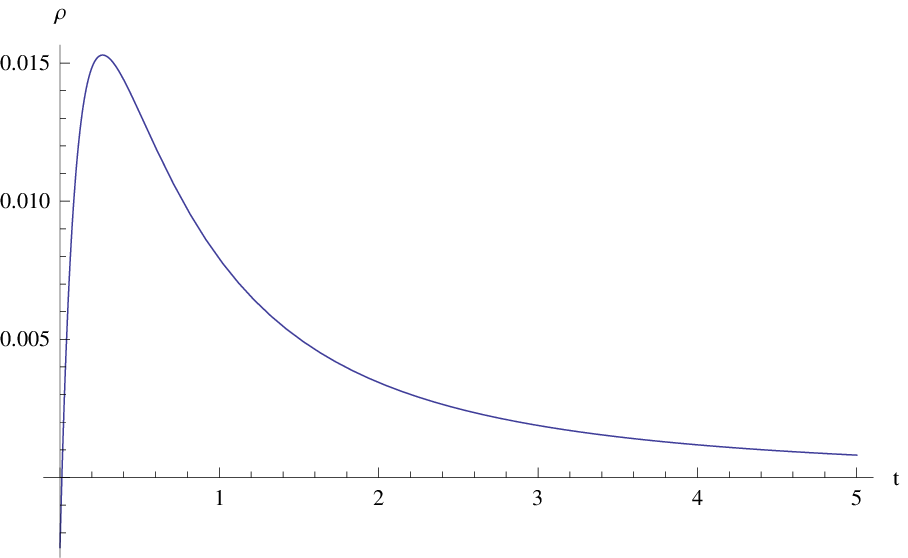,width=0.5\linewidth} &
\epsfig{file=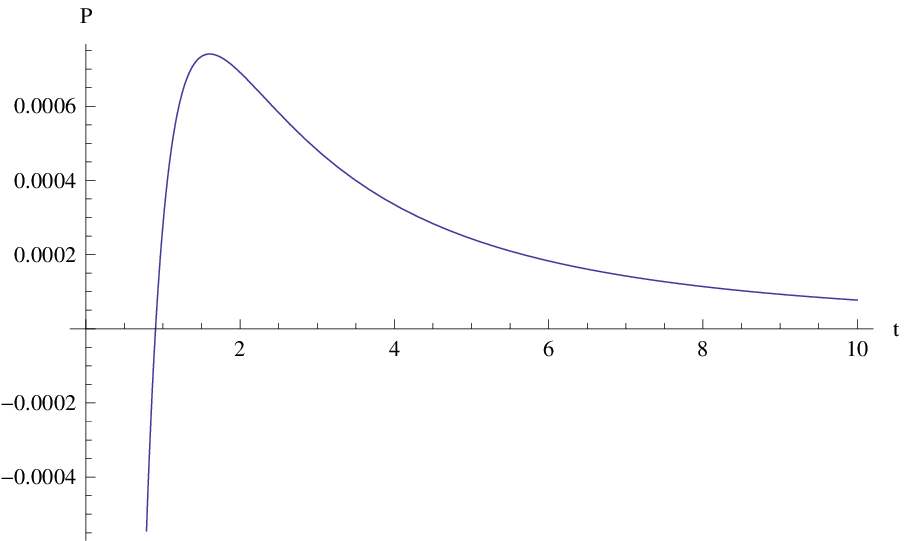,width=0.5\linewidth} \\
\end{tabular}
\caption{Behavior of energy density and pressure versus time for $t>0$ with
$n=2,~\lambda=1,~l=1,~k_1=0,~q_1=1=q_2$ and
$q_3=-2$.}\center
\end{figure}
\begin{figure}
\begin{center}
\includegraphics[width=2.8in, height=1.8in]{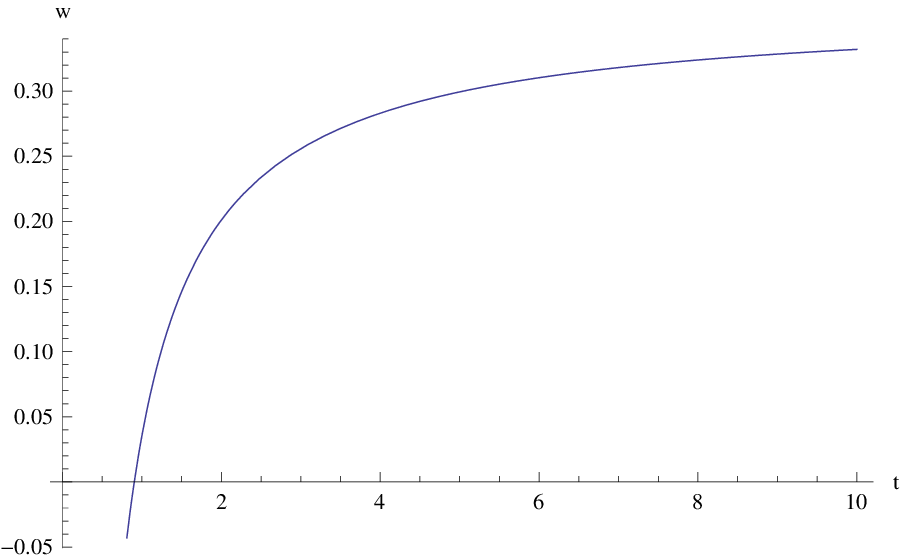}
\caption{Behavior of $w$ versus time for $t>0$ with
$n=2,~\lambda=1,~l=1,~k_1=0,~q_1=1=q_2$ and
$q_3=-2$.}\center
\end{center}
\end{figure}

\subsection{Non-singular Model of the Universe}

For this model, $n=0$ and the average scale factor $a=k_2\exp(lt)$
turns the metric coefficients $A,~B$ and $C$ into
\begin{eqnarray}\label{43}
A&=&p_1k_2\exp(lt)\exp\bigg[-\frac{q_1\exp(-3lt)}{3l{k_2}^3}\bigg],\\\label{44}
B&=&p_2k_2\exp(lt)\exp\bigg[-\frac{q_2\exp(-3lt)}{3l{k_2}^3}\bigg],\\\label{37}
C&=&p_3k_2\exp(lt)\exp\bigg[-\frac{q_3\exp(-3lt)}{3l{k_2}^3}\bigg].
\end{eqnarray}
The directional Hubble parameters $H_i$ become
\begin{equation}\label{44}
H_i=l+\frac{q_i}{{k_2}^3}\exp(-3lt) .
\end{equation}
The mean generalized Hubble parameter and volume scale factor
turn out to be
\begin{equation}\label{46}
H=l,\quad V={k_2}^3\exp(3lt).
\end{equation}
\begin{figure}\center
\begin{tabular}{cccc}
\epsfig{file=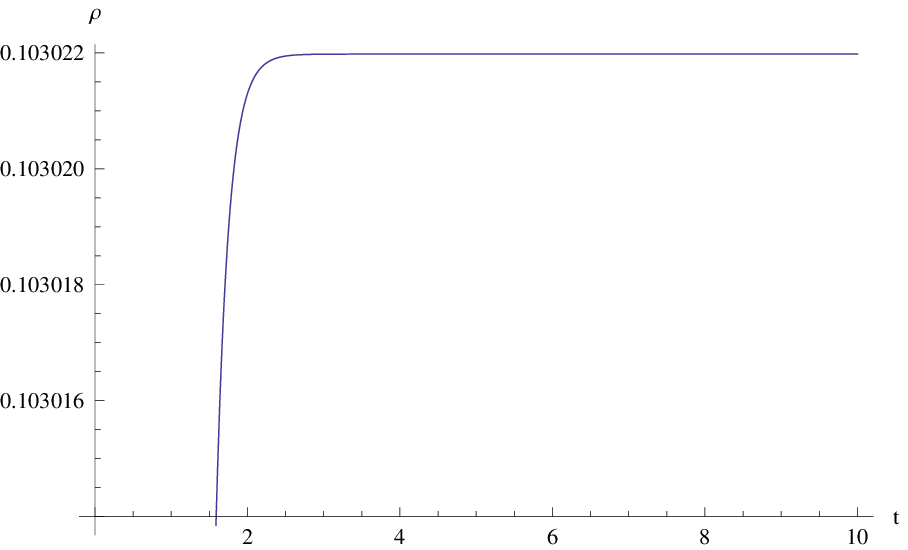,width=0.5\linewidth} &
\epsfig{file=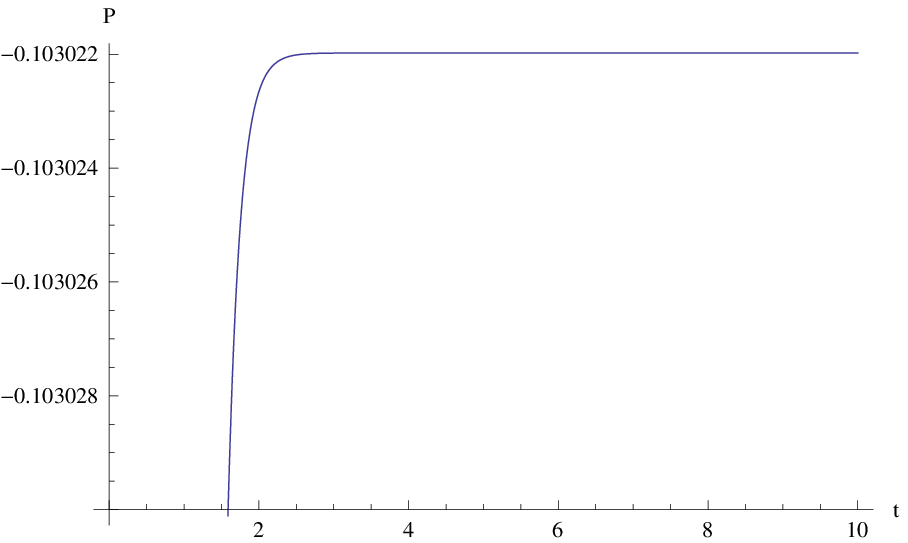,width=0.5\linewidth} \\
\end{tabular}
\caption{Behavior of energy density and pressure versus time for $t>0$ with
$n=2,~\lambda=1,~l=1,~k_2=1,~q_1=1=q_2$ and
$q_3=-2$.}\center
\end{figure}
The mean anisotropy parameter, expansion scalar and shear scalar
are
\begin{equation}
A=\frac{{q_1}^2+{q_2}^2+{q_3}^2}{3l^2{k_2}^6\exp(6lt)},~~
\theta=3l,\quad
\sigma^2=\frac{{q_1}^2+{q_2}^2+{q_3}^2}{2{k_2}^6\exp(6lt)}.
\end{equation}
The energy density and pressure of the universe
take the form
\begin{eqnarray}\nonumber
\rho=&&\frac{1}{12(\lambda+2\pi)(\lambda+4\pi)}\bigg[4(\lambda+3\pi)
\bigg\{3l^2+\frac{q_1q_2+q_2q_3+q_3q_1}{{k_2}^6\exp(6lt)}\bigg\}\\\label{ro2}
&-&\lambda\bigg\{3l^2+\frac{{q_1}^2+{q_2}^2+{q_3}^2}{{k_2}^6\exp(6lt)}\bigg\}\bigg],
\end{eqnarray}
\begin{eqnarray}\nonumber
p=&&\frac{-1}{12(\lambda+2\pi)(\lambda+4\pi)}\bigg[4\pi
\bigg\{3l^2+\frac{q_1q_2+q_2q_3+q_3q_1}{{k_2}^6\exp(6lt)}\bigg\}\\\label{p2}
&+&(3\lambda+8\pi)\bigg\{3l^2+\frac{{q_1}^2+{q_2}^2+{q_3}^2}{{k_2}^6\exp(6lt)}\bigg\}\bigg].
\end{eqnarray}
For this model, the plots of $\rho$, $P$ and $w$ against
time coordinate $t$ are shown in figure $3$ and $4$ respectively.
It can be seen from figure $4$ that $w\rightarrow -1$ as $t\rightarrow\infty$
which indicates that the non-singular model corresponds to a vacuum
fluid dominated universe.
\begin{figure}
\begin{center}
\includegraphics[width=2.8in, height=1.8in]{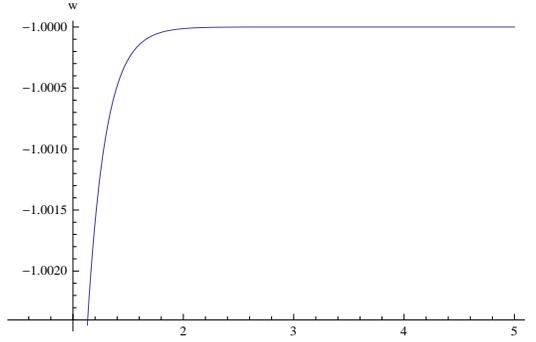}
\caption{Behavior of $w$ versus time for $t>0$ with
$~n=2,~\lambda=1,~l=1,~k_2=1,~q_1=1=q_2$ and
$q_3=-2$.}
\end{center}
\end{figure}

\section{Concluding Remarks}

This paper is devoted to discuss the current phenomenon of
accelerated expansion of universe in the framework of newly
proposed $f(R,T)$ theory of gravity. For this purpose, we take
$f(R,T)=R+2\lambda T$ and explore the exact solutions of Bianchi
type $I$ cosmological model. We obtain two exact solutions using
the assumption of constant value of deceleration parameter and the
law of variation of Hubble parameter. The obtained solutions
correspond to two different models of universe. The first solution
forms a singular model with power law expansion while the second
solution gives a non-singular model with exponential expansion of
universe. The physical parameters for both of these models are
discussed below.

The singular model of the universe corresponds to $n\neq0$ with
average scale factor $a=(nlt+k_1)^{\frac{1}{n}}$. This model
possesses a point singularity when $t\equiv t_s=-\frac{k_1}{nl}$.
The volume scale factor and the metric coefficients
$A,~B$ and $C$ vanish at this singularity point. The cosmological
parameters $H_1,~H_2,~H_3,~H,~ \theta$, and $\sigma^2$ are all
infinite at this point of singularity. If we choose $k_1=0$, figure $1$ suggests that
energy density of the universe is zero at this time. The pressure
approaches negative infinity as $t\rightarrow 0$. This strong negative pressure is an
indication of dark energy. For this model, $w\rightarrow \frac{1}{3}$ as $t\rightarrow\infty$ which
corresponds to a radiation dominated universe.
The mean anisotropy parameter $A$ also
becomes infinite at this point for $0<n<3$ and vanishes for $n>3$.
Moreover, the isotropy condition, i.e.,
$\frac{\sigma^2}{\theta}\rightarrow 0$ as $t\rightarrow \infty$,
is verified for this model. All these conclusive observations
suggest that the universe starts its expansion with zero volume,
strong negative pressure and energy density from $t=t_s$ and it will
continue to expand for $0<n<3$.

Now we discuss the non-singular model of the universe corresponds
to $n=0$. For this model the average scale factor is
$a=k_2\exp(lt)$. The non-singularity is due to the exponential
behavior of the model. The expansion scalar $\theta$ and mean
generalized Hubble parameter $H$ are constant in this case. For
finite values of $t$, the physical parameters
$H_1,~H_2,~H_3,~\sigma^2$ and $A$ are all finite. The metric
functions are defined for finite time and the isotropy condition
is satisfied. There is an exponential increase in the volume as the time grows.
However, energy density is approximately zero initially and
becomes constant after some time. Pressure of the universe remains
in the negative zone for this model which may be an indication of
presence of dark energy in the universe. Figure $4$ suggests
that $w\rightarrow -1$ as $t\rightarrow\infty$. Thus the exponential model
corresponds to a vacuum fluid dominated universe.
According to the observations \cite{nature}, the expansion of the universe is accelerating when $w\approx-1$.

Therefore, it is hoped that the problematic issues such as dark energy
and accelerated expansion of universe may be addressed
using modified theories of gravity especially $f(R,
T)$ gravity. It would be interesting to explore more Bianchi type
solutions in this context. Exact solutions of Bianchi type $V$
cosmological model in this theory are under process.\\\\
\textbf{Acknowledgement}\\\\ The author is thankful to National University
of Computer and Emerging Sciences (NUCES) Lahore Campus for
funding the PhD programme. The author is also grateful to the anonymous reviewer
for valuable comments and suggestions to improve the paper.

\end{document}